\documentstyle[multicol,aps,prb,epsf]{revtex}
\begin{document}

\draft

\title{
Image-potential band-gap narrowing at a metal/semiconductor interface
}

\author{
Ryotaro Arita, Yoshiaki Tanida$^1$, Kazuhiko Kuroki$^2$, and Hideo Aoki 
}

\address{Department of Physics, University of Tokyo, Hongo,
Tokyo 113-0033, Japan}
\address{$^1$Fujitsu Laboratories Ltd.,10-1, Morinosato-Wakamiya, 
Atsugi, Kanagawa 243-0197, Japan}
\address{$^2$Department of Applied Physics and Chemistry,
University of Electro-Communications, Chofu, Tokyo 182-8585, Japan}

\date{\today}

\maketitle

\begin{abstract}
GW approximation is used to systematically revisit 
the image-potential band-gap narrowing at 
metal/semiconductor interfaces proposed by Inkson in the 1970's.  
Here we have questioned how the narrowing as 
calculated from quasi-particle energy spectra for the
jellium/Si interface depends on  $r_s$ of the jellium.
The gap narrowing is found to only weakly 
depend on $r_s$ (i.e., narrowing $\simeq 0.3$ eV 
even for a large $r_s = 6)$.  Hence we can turn to 
smaller polarizability in the semiconductor side as an 
important factor in looking for larger narrowing.
\end{abstract}

\medskip


\begin{multicols}{2}
\narrowtext

\section{Introduction}
Disrupted translational symmetry at surfaces and interfaces 
provides a potentially rich playing ground 
for many-body effects.  A seminal proposal was 
in fact made by Inkson back in the 1970's, who proposed that 
a metal-insulator transition can take place around the 
interface.\cite{Inkson1971a,Inkson1971b,Inkson1972,Inkson1973,Anderson1974}
Classically, his argument is as follows.
When a metal is placed on top of a 
semiconductor, an electron in the semiconductor feels an 
image potential which is the interaction between the particle and 
its image charge in the metal.
This leads to a downward bending of the conduction band bottom 
toward the interface, 
$1/(4\varepsilon z$), where $\varepsilon$ is the dielectric constant of
the semiconductor and $z$ is the distance from 
the interface.
Similarly, the valence band top is bent upward by the same amount.

Quantum mechanically, this becomes as follows\cite{Inkson1973}.
The contribution of the correlation term to the self-energy
of the electron in the semiconductor
is similar ($\sim -1/(4\varepsilon z)$)
for the conduction and valence bands.  
On the other hand, while the screened exchange term almost 
vanishes for the 
conduction band, it amounts to ($\sim 1/(2\varepsilon z)$)
for the valence band.  As a result, while
the conduction band bends downward like $-1/(4\varepsilon z$),
the valence band bends upward like $1/(4\varepsilon z$).  
This kind of band bending can occur over 
short distances ($z \sim O(10)$ \AA), 
while the usual Schottky barrier occurs 
over much larger distances ($z \sim O(100-1000)$ \AA).

After the proposal of Inkson was made, various studies for 
the band gap reduction or closure 
at metal/semiconductor(insulator) interfaces have been performed
theoretically and experimently\cite{Murata2000,Kiguchi2001,Okiji1979,Charlesworth1993}.
Recently, Murata {\it et al}\cite{Murata2000} 
have studied Ru(0001)/Al$_2$O$_3$ and have observed the
band gap narrowing of Al$_2$O$_3$.
Kiguchi {\it et al}\cite{Kiguchi2001} 
have studied LiCl films on Cu(001), and have found
that 3p level of Cl shift up to the Fermi level
as the number of LiCl layers decreases.

As for the first principles many body calculation,
Charlesworth {\it et al.}\cite{Charlesworth1993} 
have calculated the quasiparticle electronic
structure of Al/GaAs(110), and have  shown 
the band gap narrowing for the first time.
The amount of the band gap reduction turned out to be about 0.4 eV.

A big issue remains however: 
which combination of metal and semiconductor will 
favor the local metal-insulator transition? 
As a first step toward such studies,
we should investigate how the band gap reduction 
depends on the density of electrons (as represented 
by $r_s$) in the metallic side, which governs 
the electron correlation in that side.  
The value of $r_s$ in metals in fact extends over a wide range, 
1.8 $\sim$ 5.6, 
and the image-potential effect in metals with 
greater $r_s$ is expected to be smaller 
than that for a smaller $r_s=2.1$, a value corresponding to 
Al and assumed by Charlesworth\cite{Charlesworth1993}.
If the band gap narrowing still occurs significantly for
metals with larger $r_s$, we can turn our attention 
to the semiconductor side in optimizing 
the local metal-insulator transition.

This is exactly the purpose of the present paper, i.e., 
to discuss the $r_s$ dependence of the band gap reduction 
quantitively.  
Since $r_s$ governs the dielectric response, the image-potential
effect may well depend sensitively on $r_s$, which is why
we have to look into the dependence from first principles.
For that purpose we have to go beyond the usual 
local density functional approximation (LDA), since we are 
talking about the effects of screening.  So here we adopt 
what is called the GW approximation, which is 
roughly the RPA (random-phase approximation) + LDA.  

\section{Model}

To focus on the problem described above, we can simplify the 
metallic side into the jellium model.  On the other hand, 
we have to have an atomistic model for the semiconductor 
side, since we are questioning effects occurring on the 
length scale of few atomic spacings on this side.  
So the model is depicted in Fig.1.  To 
facilitate the band calculation, we 
adopt a repeated-slab model (periodic boundary condition), in which 
the semiconductor slabs alternate with the 
jellium slabs.  
We calculate the band structure in the $k_x$-$k_y$ space.

When a semiconductor and a metal are put together, 
the Fermi energies in the metal and semiconductor have to be 
aligned in thermal equilibrium, which implies that  
some charges should flow across the interface 
resulting in a charged region on the semiconductor side 
in general (in the absence of impurity levels in the
semiconductor gap).  
This charge redistribution creates an electrostatic 
potential, and this bends the valence and conduction bands
of the semiconductor, which becomes significant at distances 
$z\sim O(100\sim 1000)$ \AA.
Since what we want to look at is the physics on 
the scale of $O(10\sim 100)$ \AA,
we can neglect this bending 
(unless the charge rearrangement is drastic).
In fact, Charlesworth {\it et al.}\cite{Charlesworth1993} 
have estimated the effect of the charge redistribution
for Al/GaAs(110), where electrons flow from Al with a higher Fermi energy
to the lower conduction band bottom in GaAs,
and found the effect to be negligible.
However, when the charge rearrangement occurs drastically,
this one-body effect can become serious.
Thus, to focus on the image-potential
effect we have to exclude the one-body potential effect carefully.
For this purpose, we set the Fermi energy of the metal
inside the energy gap of the semiconductor(Fig.2).
As we shall show, such a situation is indeed realized if we take an appropriate
value of $r_s$ for the jellium model. 

As for the semiconductor side, we have employed a slab, 
which consists of five layers of Si atoms 
stacked in the [001] direction. 
To get rid of complications arising from dangling bonds,
the edges are terminated by hydrogen atoms.
For the structure of the hydrogen-terminated surface, 
we assume for simplicity a non-reconstructed one
($p(1\times 1)$). With this assumption we have optimized 
the structure imposing the
mirror-plane symmetries (along [100] and [010]) and an 
inversion+mirror symmetry ([001]). 
The hydrogen atoms are  allowed to relax in any directions.
The size of a supercell 
is 7.18$^2$ $\times$ 30.48 a.u.$^3$ along the $(x,y,z)$, with 
the thickness of the jellium being 12.51 a.u. 

The band gap narrowing or closure in the semiconductor is 
probed here by identifying the character of the wave functions 
for various bands for the repeated slab model: by concentrating 
on the bands whose wave functions have their amplitudes primarily on the 
semiconducting side we can define the gap of the semiconductor.  

\section{Method}
Band-structure calculations are usually 
performed within the framework of the LDA.
In this formalism, many-body effects 
are represented by the so-called exchange-correlation potential, 
which is a functional of the electron density.
In practice, this potential is  
approximated as a function of the local density, 
and we have the LDA.

While the density-functional formalism is shown to 
be rigorous for 
the ground state\cite{Hohenberg1964}
and  LDA gives reliable information about the
ground state properties for various electron systems,
it is well-known that these approaches are 
not useful for excited states.
In fact, LDA usually underestimates
the band gap of semiconductors and insulators.
Moreover, LDA cannot be applied to the cases 
where the electron density varies in space.

Still, LDA wave functions are usually
good approximation to quasi-particle wave functions\cite{Aulbur2000}.
Since excitations can be described by many-body perturbation theory,
it should be quite a good starting point to adopt the 
LDA wave functions as the basis for the many-body perturbation theory 
in determining the self-energy and spectrum of the quasi-particles. 

For the calculation of the self-energy, 
various approximations have been developed.
Among them, Hedin's GW approximation\cite{Hedin1965,Hedin1969} often
gives excellent quasi-particle energies in
bulk semiconductors with a comparatively 
simple formalism\cite{Aulbur2000,Aryasetiawan1998}.
The GW approximation essentially amounts to the RPA 
in the LDA formalism, so we have adopted this method to 
study the image-potential band-gap reduction.

\subsection{LDA}
So the first task in the present study is to perform 
an LDA calculation to obtain the eigenwavefunctions
for the system described above. 
We adopt the exchange-correlation functional introduced by
Perdew and Wang\cite{Perdew1992}, and
eigenwavefunctions are expanded by plane waves up to
a cut-off energy of 16Ry.
As for the atomic pseudo-potentials, 
soft, norm-conserving pseudo-potentials
in a separable form\cite{Troullier1991} are employed.
The atomic configurations and the corresponding electronic states
in the ground states are obtained with the conjugate gradient
scheme\cite{Yamauchi1996}.

\subsection{Self-energy correction in the GWA}
We then proceed to the GW approximation (GWA)calculation.
The central idea of GWA  is 
to approximate the self-energy operator $\Sigma$ by
\[
\Sigma({\bf r},{\bf r'};\omega)=\frac{i}{2\pi}
\int d\omega' 
G({\bf r},{\bf r'};\omega+\omega')
W({\bf r},{\bf r'};\omega ')
e^{i\omega' \delta},
\]
where $\delta$ is an infinitesimal positive time
and $W$ is the screened Coulomb interaction,
\[
W({\bf r},{\bf r'};\omega)=\int d{\bf r''}
\frac{1}{\varepsilon({\bf r''},{\bf r'}; \omega)|{\bf r}-{\bf r''}|},
\]
where $\varepsilon$ is the dielectric function.

Recently, Rojas {\it et al}\cite{Rojas1995} proposed
a new implementation of GWA, the space-time approach,
which is described in detail by Rieger {\it et al}\cite{Rieger1999}.
The quasi-particle calculations can be performed 
either in reciprocal space as a function of the 
frequency, or in real space as a function of the imaginary time,
and the central idea in this method is to 
choose the representation that minimizes 
the computations required to evaluate the basic GWA quantities.
This approach enables us to study larger systems.

The actual computational steps in this method are as follows.
First, split the self-energy into a bare exchange part $\Sigma^X$
and an energy-dependent correlation contribution $\Sigma^C(E)$.
The former can be calculated from
\begin{eqnarray*}
\langle m {\bf k} | \Sigma^X |m {\bf k} \rangle =
-\frac{4\pi}{V}\sum_{v}^{\rm occ}\sum_{{\bf q,G}}
\frac{|M_{\bf G}^{vm}({\bf k},{\bf q})|^2}{|{\bf q}+{\bf G}|^2}.
\end{eqnarray*}
Here ${\bf G}$ is the reciprocal vector, and 
\begin{eqnarray*}
M_{\bf G}^{vm}({\bf k},{\bf q})=\int
\Phi_{v,{\bf k-q}}({\bf r})
e^{-i({\bf q+G})\cdot {\bf r}}
\Phi_{m,{\bf k}}({\bf r})d{\bf r},
\end{eqnarray*}
where $\Phi_{v,{\bf k-q}}$ is the wave function in the 
valence band of the semiconductor, while $\Phi_{m,{\bf k}}$ 
is the wave function in the $m$-th band.

To evaluate the energy-dependent self-energy $\Sigma^C(E)$,
we first construct Green's function in 
real space and imaginary time,
\[G_{LDA}= \cases{\displaystyle{
i\sum_{n{\bf k}}^{\rm occ} \Phi_{n{\bf k}}({\bf r})
\Phi_{n{\bf k}}^*({\bf r})\exp(\epsilon_{n{\bf k}}\tau)} ,\tau >0 \cr
\displaystyle{ 
-i\sum_{n{\bf k}}^{\rm unocc}\Phi_{n{\bf k}}({\bf r})
\Phi_{n{\bf k}}^*({\bf r})\exp(\epsilon_{n{\bf k}}\tau)}  ,\tau <0 \cr
}\]
where $\Phi_{n {\bf}}$ and $\epsilon_{n {\bf}}$ 
are LDA wavefunctions and eigenvalues. 
Due to a rapid decay of the exponentials, the convergence 
against the cut-off in $\sum_{n{\bf k}}^{\rm unocc}$ is much better than 
that in the real frequency formalism.

Next, the RPA irreducible polarizability 
$\chi^0({\bf r},{\bf r'};i\tau)$ is
calculated in real space and imaginary time, and Fourier-transformed
to reciprocal space and imaginary energy.
Then, the symmetrized Hermitian dielectric matrix\cite{Baroni1986} 
$\tilde{\varepsilon}_{{\bf G},{\bf G'}}({\bf k},i\omega)$
is constructed, and inverted for each {\bf k} point.

Then the screened Coulomb interaction is calculated as
\begin{eqnarray*}
W_{{\bf G}{\bf G'}}({\bf k},i\omega)=
\frac{4\pi}{|{\bf k+G}||{\bf k+G'}|}
\tilde{\varepsilon}_{{\bf G},{\bf G'}}({\bf k},i\omega)^{-1},
\end{eqnarray*}
and is Fourier-transformed to real space and imaginary time.
The self-energy operator can be calculated as
\[
\Sigma^C({\bf r},{\bf r'};i\tau)=iG_{LDA}({\bf r},{\bf r'};i\tau)
W({\bf r},{\bf r'};i\tau).
\]
Finally, we evaluate the correlation contribution, 
$\Sigma^C(i\tau)= \langle {\bf k}n |\Sigma^C|{\bf k}n \rangle$.
After this is Fourier-transformed to imaginary energy,
we perform analytic continuation onto the real energy axis 
with the Pad\'{e} approximation\cite{Vidberg1977}.

In the present study, we consider 
a $6 \times 6 \times 24$ grid for the unit cell, and
a $6 \times 6$ grid for $(k_x,k_y)$.  
The time grid is spaced by $\delta \tau = 0.3$ a.u.
over the range of 13 a.u.
We have taken up to 253 states to construct Green's function.

\section{Result}
Let us move on to the results. 
The quasi-particle spectrum $\epsilon^{QP}$ is obtained as 
\[
\epsilon^{QP}=\epsilon^{LDA}-V_{xc}^{LDA}+\Sigma^X+\Sigma^C.
\]
In the following, 
we discuss the effect of the metallic layer 
for each term in $\epsilon^{QP}$.

\subsection{The LDA calculation}
We first show the band structure and the
(squared absolute value of) the wave functions
obtained with LDA for $r_s=\infty$, i.e., no jellium (Fig.3),
$r_s=6$ (Fig.4) and $r_s=4$ (Fig.5), respectively.
In the absence of the metal ($r_s=\infty$),
we can see that the valence band top is at
$\Gamma$, while the conduction band bottom lies around 
$K$ (and $\Gamma$).
Hereafter, we focus on the energy shifts 
of the valence band top around $\Gamma$ and
the conduction band bottom around $K$ 
caused by the close contact with the jellium.

When in vacuum 
the LDA band gap across $K$ and $\Gamma$ is 1.93 eV.
When we introduce the jellium with $r_s=6$ or 4, 
the Fermi level still lies across $K$ and $\Gamma$, i.e.,
only a small amount of electrons flow from the jellium into the Si side,  
so the effect of the charge rearrangement is almost completely absent.
For smaller values of $r_s$, on the other hand, 
we can show that electrons in the jellium do 
flow into the Si side, and the electronic 
band structure changes drastically.
Thus  we focus here on the case of $r_s=6$ and 4.

For $r_s=6 (4)$, the band gap 
across $\Gamma$ and K reduces to 1.85 (1.79) eV.
The characters of the wave functions at these points 
are mainly Si and do not change, as we can see in Fig.3, 4, and 5. 
In addition we notice that there is a state which emerges around 
$\Gamma$ crossing the Fermi level for $r_s=4,6$ (the bands 
represented by bold lines in the figures).  If we examine 
the character of wave functions on these branches in Figs. 4,5, 
they reside well within the metallic side, so 
we exclude them from our argument on the band-gap 
narrowing in the semiconductor side.  

\subsection{$V_{XC}$ and the direct exchange term $\Sigma^X$}
Now we are in position to analyse the band gap reduction term by term.  
Here we consider
$V_{XC}$ and the direct exchange term $\Sigma^X$, 
the energy independent part of the GW correction
to the LDA eigenenergy.
The expectation value of the 
exchange-correlation potential is obtained by
$\langle V_{\rm xc}^{LDA} \rangle \equiv
\int |\Phi_{{\bf k},n}({\bf r})|^2 V_{\rm xc}^{LDA}
({\bf r}) d{\bf r}$.
We have found that $V_{\rm xc}({\bf r})^{LDA}$ in the Si region
has similar values for $r_s =\infty$, 6 and 4.
Since the characters of the wave functions
do not change as we have seen in Figs.3, 4, and 5,
we can expect that
$\langle V_{\rm xc}^{\rm LDA} \rangle$ does not change significantly 
when we introduce the jellium between the Si system.

On the other hand, 
the value of the matrix elements between 
the metal wave functions and the semiconductor 
wave functions, 
$M_{\bf G}^{vc}({\bf k},{\bf q})$, which governs $\Sigma^X$, 
are small, since the states of jellium character and
Si character are well separated in real space (see Figs.4,5).  
Therefore, we can expect that $\Sigma^X$ does not 
change significantly when we introduce the jellium.

In fact, $-V_{XC}+\Sigma^X$ at 
K is $3.99 \rightarrow 4.00 \rightarrow 3.98$ in eV 
for $r_s=\infty\rightarrow 6 \rightarrow 4$, 
so the shift is negligible.
At $\Gamma$, $-V_{XC}+\Sigma^X$ is 
$-2.09  \rightarrow -2.00 \rightarrow -1.93$ 
in eV for $r_s=\infty  \rightarrow 6 \rightarrow 4$.  

\subsection{Energy-dependent $\Sigma^C$}
Finally, let us discuss the correlation contribution, $\Sigma^C$.
In Fig. 6, we show the imaginary frequency dependence of
$\Sigma^C$ at $\Gamma$ and K.

We have then performed an analytic continuation onto the real energy
axis with the Pad\'{e} approximation\cite{Vidberg1977}.
In Fig.7, we show the real frequency dependence of 
$\Sigma^C$. We can see that there is 
an almost linear dependence on $\omega$.

When $\Sigma^C$ is linear for small $\omega$, 
the GW correction to the LDA spectrum 
reduces to
\[
\Delta=\frac{1}{Z_{n {\bf k}}}\langle
\Phi_{n {\bf k}} |
\Sigma^C(\epsilon_{n {\bf k}}^{LDA})+\Sigma^X-V_{XC}
|\Phi_{n {\bf k}} 
\rangle
\]
where
\[
Z_{n {\bf k}}=1-\frac{d}{d\omega}
\langle
\Phi_{n {\bf k}} |
\Sigma^C(\omega)
|\Phi_{n {\bf k}} 
\rangle |_{\omega=\epsilon_{n {\bf k}}^{LDA}}.
\]
From Fig.7, we can see that $Z_{n {\bf k}}= -0.2 \sim -0.3$.
At $\Gamma$, $\Delta$ is estimated to be $-0.75 \rightarrow -0.60
\rightarrow -0.44$ in eV for $r_s=\infty  \rightarrow 6 \rightarrow 4$. 
At K, $\Delta$ is $1.28  \rightarrow 1.15 \rightarrow 1.09$ 
in eV.  Thus the band-gap reduction due to the presence of the 
jellium amounts to as large as 
$\approx 0.3$ eV for $r_s=6$ and $\approx 0.5$ eV
for $r_s=4$.

\section{Discussions}
To summarize, we have studied, with the GW approximation
and a character-resolved band analysis, 
the image-potential band-gap narrowing at a
metal/semiconductor interface
by calculating quasi-particle energy spectrum of the
jellium/Si interface.  
For the values of $r_s=4 - 6$ of jellium studied here 
the electrons or holes do not flow from the metallic to 
semiconducting side, i.e., the Fermi energy of the 
jellium lies within  the energy gap of Si, so 
the one-body effect due to the
charge redistribution is absent.
We have found that a significant  
band gap narrowing of $\approx 0.3$ eV occurs for $r_s$ 
as large as $6$.

So we can concentrate on the semiconducting side to 
realize larger gap-narrowing effects or a local 
metal-insulator transition.  
If the dielectric constant of the semiconductor
is small, 
the image potential effect will become stronger, 
so that we may expect larger band gap narrowing.
However, the system with a small dielectric function
usually occurs in materials with large 
band gap, so that the realization of the band-gap 
closure becomes a trade-off. 
Furthermore, the energy gap of the semiconducting layer
may depend on the surface structure 
(e.g., whether it is terminated by 
hydrogen atoms or dangling bonds form dimers, etc.), 
and their effect is also non-trivial.
These are interesting future problems.

Finally, while
the band gap narrowing or closure is of a fundamental 
interest in its own right,  we can further raise 
a very strong motivation for considering a metallized 
semiconductor surface.  
There is a long history\cite{Ginzburg82} of proposals for 
superconductivity in conducting systems in close contact with 
polarizable media.  Little\cite{Little64} 
proposed this for one-dimensional 
systems, then Ginzburg\cite{Ginzburg64} extended this to two-dimensional 
systems. Allender, Brey and Bardeen\cite{Allender73}
studied this in detail, 
with metal-semiconductor structures in mind, which was 
subsequently criticized by Inkson and Anderson\cite{InksonAnderson73}.  
A summary of the situation by Zharkov\cite{Ginzburg82} 
is that, while the criticism is correct for the usual form for 
the dielectric function, the situation may be 
resurrected for unusual dielectric functions. 
The background to all this is that for the polarizable-medium 
mediated superconductivity, the conduction layer should be 
very strongly coupled to the semiconducting layer, 
ideally with strong chemical bonds such as covalent ones.  So the 
metallized semiconductor surface, with the band-closure 
mechanism, should be one ideal realization of this.  

\section{Acknowledgment}
One of the authors (KK) in indebted to Y. Murata for arousing his 
interests in the surface.   HA wishes to thank J.C. Inkson 
for illuminating discussions.  Thanks are also due to 
M. Tsukada, A. Koma and M. Kiguchi for discussions.  
The LDA calculation was performed with TAPP (Tokyo Ab-inito 
Program Package), where RA would like to thank 
S. Koizumi for technical advices. 
The numerical calculation was performed with 
SR8000 in ISSP, University of Tokyo.
This work is in part funded by a Grant-in-Aid for Scientific
Research from Ministry of Education of Japan.

\begin{figure}
\begin{center}
\leavevmode\epsfysize=50mm \epsfbox{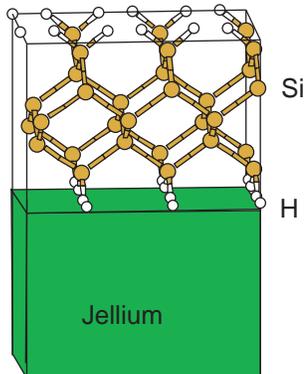}
\caption{The model studied in the present work. The slabs,
each of which consists of Si layers terminated by
H atoms, sandwich the jellium (metallic) region.}
\label{model}
\end{center}
\end{figure}

\begin{figure}
\begin{center}
\leavevmode\epsfysize=50mm \epsfbox{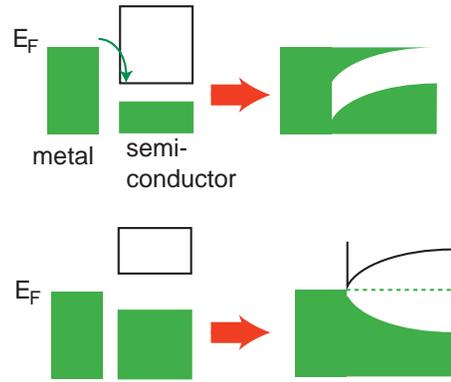}
\caption{Two cases where there is a charge transfer
across the metal-semiconductor interface (top panel)
or there is no charge redistributions with the Fermi
energy of the metal lying within the gap of the
semiconductor (bottom).}
\label{model2}
\end{center}
\end{figure}

\begin{figure}
\begin{center}
\leavevmode\epsfysize=50mm \epsfbox{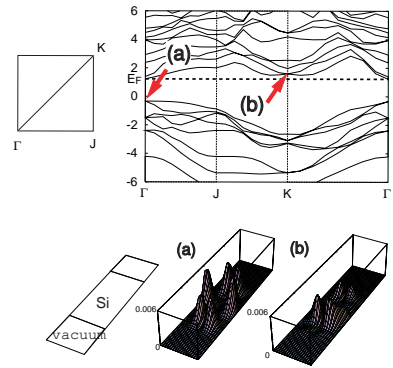}
\caption{The band structure of five Si layers 
terminated by hydrogen atoms in vacuum.
The squared absolute value of the LDA wavefunctions
at the valence top and conduction bottom are also shown.
}
\label{12-band-inf}
\end{center}
\end{figure}

\begin{figure}
\begin{center}
\leavevmode\epsfysize=50mm \epsfbox{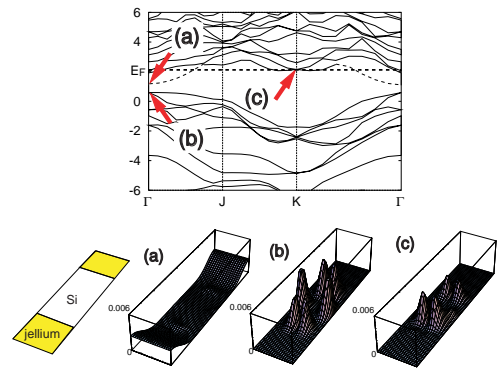}
\caption{A plot similar to Fig.3, when the 
jellium with $r_s=6$ is attached to the Si layer.
The branch whose amplitude is 
localized in the jellium region is denoted by
dotted lines.}
\label{12-band-6}
\end{center}
\end{figure}

\begin{figure}
\begin{center}
\leavevmode\epsfysize=50mm \epsfbox{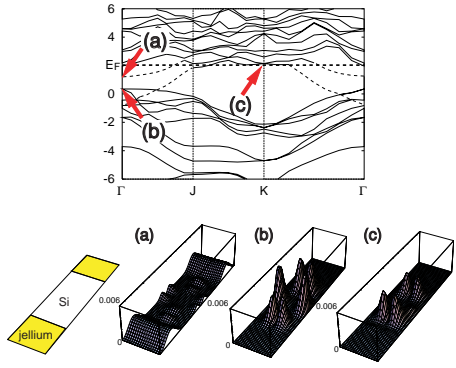}
\caption{A plot similar to Fig.4, for $r_s=4$.}
\label{12-band-4}
\end{center}
\end{figure}

\begin{figure}
\begin{center}
\leavevmode\epsfysize=60mm \epsfbox{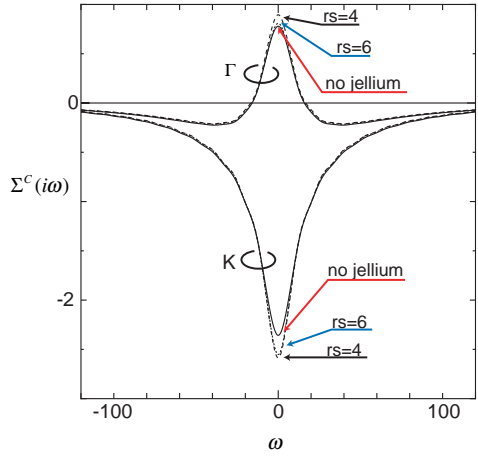}
\caption{The imaginary frequency dependence of the
energy-dependent $\Sigma^C$ at $\Gamma$ and K.
The dotted line is for $r_s=4$, the dashed line is for
$r_s=6$, and the solid line is for $r_s=\infty$.}
\label{selfenergy12}
\end{center}
\end{figure}

\begin{figure}
\begin{center}
\leavevmode\epsfysize=60mm \epsfbox{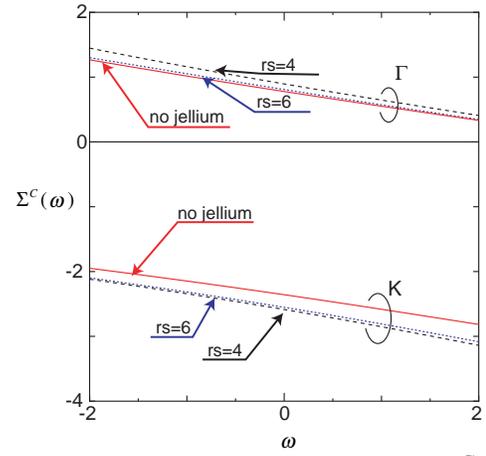}
\caption{The real frequency dependence of the
$\Sigma^C$ at $\Gamma$ and K.
The dotted line is for $r_s=4$, the dashed line is for
$r_s=6$ and the solid line is for $r_s=\infty$.}
\label{pade12}
\end{center}
\end{figure}

\end{multicols}
\end{document}